\documentclass[a4paper]{article}
\usepackage{indentfirst}
\usepackage{hyperref}
\usepackage{multirow}
\usepackage{latexsym,bm,amsmath,amssymb}
\usepackage{graphicx}
\usepackage{epsfig}
\usepackage{subfigure}
\usepackage{cite}
\usepackage{color}
\usepackage[top=1in, bottom=1in, left=1.25in, right=1.25in]{geometry}
\usepackage{slashed}
\usepackage{fancyhdr}
\usepackage{slashed}
\usepackage{makecell,rotating,multirow,diagbox}
\usepackage{tikz}
\usepackage{float}
\usepackage{appendix}
\usepackage{setspace}
\usepackage{geometry}
\usepackage{graphics}
\usepackage{multirow}
\usetikzlibrary{shapes.geometric, arrows}
\usetikzlibrary{trees}
\usetikzlibrary{decorations.pathmorphing}
\usetikzlibrary{decorations.markings}
\tikzset{
        photon/.style={decorate, decoration={snake}, draw=red},
        nucleon/.style={draw=black, postaction={decorate},
           decoration={markings,mark=at position .55 with{\arrow[draw=black]{>}}}},
        pion/.style={draw=blue, postaction={decorate},
        decoration={markings,mark=at position .55 with{\arrow[draw=blue]{}}}},
        sigma/.style={draw=black, postaction={decorate},
        decoration={markings,mark=at position .55 with{\arrow[draw=black]{}}}},
        link/.style    = { draw=black, double = white, line width = 1.8pt, double distance = 0.8pt , postaction={decorate},decoration={markings,mark=at position .55 with{\arrow[draw=black]{>}}}},
    }

\newcommand{\be}{\begin{equation}}
\newcommand{\ee}{\end{equation}}
\newcommand{\ba}{\begin{array}{c}} \newcommand{\ea}{\end{array}}
\newcommand{\bqa}{\begin{eqnarray}}
\newcommand{\eqa}{\end{eqnarray}}

\def\bea{\arraycolsep .1em \begin{eqnarray}}
\def\eea{\end{eqnarray}}

\def\s0#1#2{\mbox{\small{$ \frac{#1}{#2} $}}}
\def\0#1#2{\frac{#1}{#2}}

\begin{document}

\setcounter{topnumber}{10}
\setcounter{totalnumber}{50}
\title{Singularities and Accumulation of  Singularities of $\pi$N Scattering amplitudes}
\maketitle

\begin{center}
{\sc
Qu-Zhi Li,$^{\dagger\,}$ \,
Han-Qing~Zheng$^{\heartsuit\,,\star\,}$}
\\
\vspace{0.5cm}
\noindent{\small{$^\dagger$ \it  Department of Physics
 Peking University, Beijing 100871, P.~R.~China}}\\
\noindent{\small{$^\heartsuit$ \it  College of Physics, Sichuan University, Chengdu, Sichuan 610065, P.~R.~China}}\\
\noindent{\small{$^\star$ \it   Collaborative Innovation Center of
Quantum Matter, Beijing, Peoples Republic of China}}
\end{center}
\begin{abstract}
It is demonstrated that for the isospin $I=1/2$ $\pi$N scattering amplitude, $T^{I=1/2}(s,t)$, $s={(m_N^2-m_\pi^2)^2}/{m_N^2}$ and $s=m_N^2+2m_\pi^2$ are two accumulation points of poles on the second sheet of complex $s$ plane, and are hence accumulation of singularities of  $T^{I=1/2}(s,t)$. For $T^{I=3/2}(s,t)$,  $s={(m_N^2-m_\pi^2)^2}/{m_N^2}$ is the accumulation point of poles on the second sheet of complex $s$ plane. The proof is valid up to all orders of chiral expansions.
\end{abstract}

In a previous publication, it is pointed out that in the $L_{2I,2J}=S_{11}$ channel partial wave $\pi$N scattering amplitude, there exist two virtual poles.~\cite{Li:2021tnt} One lies below but close to $c_L=\frac{(m_N^2-m_\pi^2)^2}{m_N^2}$, and another lies above but close to $c_R=m_N^2+2m_\pi^2$. Here $[c_L, c_R]$ defines the cut of partial wave amplitude caused by the $u$ channel nucleon exchange.~\cite{Kennedy:1962apa} Meanwhile it is found that the dispersion representation for the phase shift has to be modified, and the net effect of the new contributions from virtual poles and the additional cut is vanishingly small, leaving previous analyses on  the $S_{11}$ partial wave  unchanged.
In this note we will extend our previous analysis to all partial waves $L_{2I,2J}$. We find that the existence of virtual poles are quite universal, but different channels behaves rather differently.

The on-shell  $T$-matrix elements for  elastic $\pi$N scatterings depend on three Mandelstam variables:
\begin{equation}
s=(p+q)^2,\quad t=(p-p^\prime),\quad u=(p-q^\prime)\ ,
\end{equation}
obeying  the constraint
\begin{equation}
	s+t+u=2m_N^2+2m_\pi^2\ ,
\end{equation}
with $m_N$ and $m_\pi$ the physical masses of the nucleon and the pion, respectively. In the isospin limit,
the isospin  amplitude can be decomposed as,~\cite{Gasser:1988NP}
\begin{equation}
	T(\pi^a+N_i\to\pi^\prime +N_f)=\chi_f^\dagger(\delta^{aa^\prime}T^+ + \dfrac{1}{2}[\tau^{a^\prime},\tau^a]T^-)\chi_i\ ,
\end{equation}
where $\tau^a$ is Pauli matrices, and {$\chi_i$ ($\chi_f$) corresponds to the isospin wave function of the initial  (final) nucleon state}.
The amplitudes with isospins $I={1}/{2}, {3}/{2}$ can be written as
\begin{equation}
	\begin{split}
& T^{I=1/2}=T^+ +2T^-\ ,\\
& T^{I=3/2}=T^+-T^-\ .
	\end{split}
\end{equation}
As for the Lorentz structure, for an isospin index $I = {1}/{2}, {3}/{2}$,
\begin{equation}
	T^I =\bar u^{(s^\prime)}(p^\prime)[A^I(s,t)+\dfrac{1}{2}(\slashed q+\slashed q^\prime)]B^I(s,t)]u^{(s)}(p)\ ,
\end{equation}
where the superscripts {($s^\prime$), (s)} denote the spins of the Dirac spinors, and q(p) and $q^\prime (p^\prime)$ are the 4-momenta of initial and final pions (nucleons), respectively. Functions $A$ and $B$ are scalar functions of $s$ and $t$.
Helicity amplitudes can be obtained by substituting the nucleon spinors by helicity  eigenstates in the centre of mass  frame
\begin{equation}\label{TPM}
	\begin{split}
&T^I_{++}=(\dfrac{1+z_s}{2})^{1/2}[2m_NA^I(s,t)+(s-m_\pi^2-m_N^2)B^I(s,t)]\ ,\\
& T^I_{+-}=-(\dfrac{1-z_s}{2})^{1/2}s^{-1/2}[(s-m_\pi^2+m_N^2)A^I(s,t)+m_N(s+m_\pi^2-m_N^2)B^I(s,t)]\ .
	\end{split}
\end{equation}
The first and second subscripts of $T$ refer to the helicity of initial and final nucleon, respectively, and  {``$+,-$" are shorthands for helicity $+ 1/2$ and $-1/2$, respectively;} $z_s=\cos\theta$ with $\theta$ the scattering angle.  Partial wave amplitudes for total angular momentum $J$ can be written as
\begin{equation}\label{EQ7}
	\begin{split}
	& T^{I,J}_{++}(s)=\dfrac{1}{32\pi}\int_{-1}^{1}dz_sT^I_{++}(s,t)d^J_{1/2,1/2}(z_s)\ ,\\
	&
	T^{I,J}_{+-}(s)=\dfrac{1}{32\pi}\int_{-1}^{1}dz_sT^I_{+-}(s,t)d^J_{1/2,-1/2}(z_s)\ .	
	\end{split}
\end{equation}
where $d^J$ is the standard Wigner $d$-function. {Substituting Eq.~(\ref{TPM}) into Eq.~(\ref{EQ7})}, then using the   relation between the Wigner $d$-function and the
Legendre polynomial
\begin{equation}
\sqrt{\dfrac{1\pm z}{2}}d^J_{1/2,\pm 1/2}(z_s)=\dfrac{1}{2}\left[P_{J+1/2}(z_s)\pm P_{J-1/2}(z_s)\right]\,,
\end{equation}
one gets
\begin{equation}\label{PW}
	\begin{split}
	&T^{I,J}_{++}(s)=\dfrac{1}{64\pi}[2m_NA_C^I(s)+(s-m_\pi^2-m_N^2)B_C^I(s)]\ ,\\
	&T^{I,J}_{++}(s)=-\dfrac{1}{64\pi\sqrt{s}}[(s-m_\pi^2+m_N^2)A^I_S(s)+m_N(s+m_\pi^2-m_N^2)B^I_S(s)]\ \ ,
	\end{split}
\end{equation}
with
\be F_C^I(s) = \int_{-1}^{1}dz_sF^I(s,t)[P_{J+1/2}(z_s)+P_{J-1/2}(z_s)]\ ,\ee
\be F_S^I(s) = \int_{-1}^{1}dz_sF^I(s,t)[P_{J+1/2}(z_s)-P_{J-1/2}(z_s)]\ ,\ee
where $F$ stands for $A$ or $B$.

The parity eigenstates can be constructed by the linear combinations of helicity amplitudes
\begin{equation}\label{PE}
	T^{I,J}_{\pm}=T^{I,J}_{++}(s)\pm T^{I,J}_{+-}(s)\ ,
\end{equation}
which satisfy the  optical theorem for partial waves,
\begin{equation}
\mathrm{Im}T^{I,J}_{\pm}=\rho(s)|T_{\pm}^{I,J}|^2\ .
\end{equation}
The kinematic factor $\rho(s)$ is defined as
\begin{equation}
	\rho(s)=\dfrac{\sqrt{(s-s_L)(s-s_R)}}{s}\ ,
\end{equation}
with $s_R=(m_N+m_\pi)^2$ and $s_L=(m_N-m_\pi)^2$. In fact,  amplitudes $T^{I,J}_{\pm}$ stand for  scattering states with {orbital angular momentum $l=J\mp 1/2$ and parity $P=(-1)^{l-1}$, respectively.} The partial wave $S$-matrix is  defined by
\begin{equation}\label{SM}
	S^{I,J}_{\pm}(s) = 1+2i\rho(s)T^{I,J}_{\pm}(s)\ ,
\end{equation}
which satisfies $S^\dagger S=1$. The $T$-matrix in the second Riemann sheet is written as
\begin{equation}\label{IIRS}
	T^{\mathrm{II}}(s)=\dfrac{T(s)}{S(s)}\ .
\end{equation}

The scalar function $A^I$ and $B^I$ may be calculated by chiral perturbation theory at low energies.\footnote{In the region $[s_L,s_R]$ chiral expansions are expected to work well.} Here, we present the results at $\mathcal{O}(p^1)$ level,
\begin{equation}\label{ABF}
	\begin{split}
	& A^{1/2}(s,u)=A^{3/2}(s,u)=\dfrac{m_Ng^2}{F^2}\ ,\\
	& B^{1/2}(s,u)=\dfrac{1-g^2}{F^2}-\dfrac{3m_N^2g^2}{F^2(s-m_N^2)}-\dfrac{m_N^2g^2}{F^2}\dfrac{1}{u-m_N^2}\ ,\\
	& B^{3/2}(s,u)=\dfrac{g^2-1}{F^2}+\dfrac{2m_N^2g^2}{F^2}\dfrac{1}{u-m_N^2}\ ,
	\end{split}
\end{equation}
where $F$ and $g$ denote the pion decay constant and the axial vector coupling constant, respectively.
Here we  however only need to concern the $1/(u-m_N^2)$ term, which coefficient is immune of any chiral corrections, and its sign determines the existence of the zeros (poles) of the $S$-matrix.

Substituting Eq.~(\ref{ABF})   into   Eq.~(\ref{PW}) and using Neumann equation
\begin{equation}
	Q_l(x)=\dfrac{1}{2}\int_{-1}^{1}\dfrac{P_l(y)}{x-y}dy\ ,
\end{equation}
where $Q_l$ is the Legendre function of the second type, one can get
{\begin{equation}
\begin{split}
&	B^{1/2,J}_{C,S} = -\dfrac{m_N^2g^2}{16\pi F^2 s\rho(s)^2}[Q_{J+1/2}(y(s))\pm Q_{J-1/2}(y(s))]+\cdots\ ,\\
&   B^{3/2,J}_{C,S} =  +\dfrac{m_N^2g^2}{8\pi F^2 s\rho(s)^2}[Q_{J+1/2}(y(s))\pm Q_{J-1/2}(y(s))]+\cdots\ ,
\end{split}
\end{equation}}
 in which the neglected terms {$\cdots$} denote the parts regular at $c_L$ and $c_R$ (see discussions below), which receive chiral corrections. The definition of $y(s)$ is
\begin{equation}
	y(s)=\dfrac{2m_\pi^2s-s^2+s_Ls_R}{(s-s_L)(s-s_R)}\ .
\end{equation}
Particularly, $y(c_R)=1$, $y(c_L)=-1$.
Using the explicit expressions of $Q_l(s)$,
\begin{equation}
	\begin{split}
&	Q_0(x)=\dfrac{1}{2}\ln\dfrac{x+1}{x-1}\ ,\\
& Q_l(x)=\dfrac{1}{2}P_l(x)\ln\dfrac{x+1}{x-1}-O_{l-1}(x)\ ,
\end{split}
\end{equation}
with
\begin{equation}\label{OD}
	O_{l-1}(x)=\sum_{r=0}^{\left[\dfrac{l-1}{2}\right]}\dfrac{(2l-4r-1)}{(2r+1)(l-r)}P_{l-2r-1}(x)\ ,
\end{equation}
one  finds that the function $B^{I,J}_{C,S}(s)$ contains the term
\begin{equation}
	\ln\dfrac{y(s)+1}{y(s)-1}=\ln\dfrac{m_N^2}{s}+\ln\dfrac{s-c_L}{s-c_R} \ .
\end{equation}
This term leads to the existence of  three branch points at $s=0,c_L,c_R$ in function $B^{I,J}_{C,S}$, and then in parity eigenstate amplitudes $T^{I,J}_{\pm}$. Further, when $s$ tends to $c_R$, it's easy to find that
\begin{equation}
\begin{split}
&	s\to c_R,\quad T^{1/2,J}_{\pm}\to \frac{g^2 m_N^2 (m_N^2+2 m_\pi^2)} {16\pi  F^2 (4 m_N^2-m_\pi^2)}\ln \frac{c_R-c_L}{s-c_R}\to \infty\ , \\
&	s\to c_R,\quad T^{3/2,J}_{\pm}\to -\frac{g^2 m_N^2 (m_N^2+2 m_\pi^2)} {8\pi  F^2 (4 m_N^2-m_\pi^2)}\ln \frac{c_R-c_L}{s-c_R}\to -\infty\ .	
\end{split}
\end{equation}
The sign of the infinity is independent of the angular momentum $J$.  Things will be different, however, when $s$ tends to $c_L$. It turns out that $ T^{1/2,J}_{\pm} $ tends to $ \mp(-1)^{J+1/2}\infty$, and  $ T^{3/2,J}_{\pm}$ tends to $\pm (-1)^{J+1/2}\infty$. Taking the definition  of  S-matrix Eq.~(\ref{SM}), and the fact that $2i\rho(s)$ is negative at $s=c_R,c_L$, conclusions can be drawn that:\footnote{The emergence of $S$ matrix zeros due to kinematical reasons is known, and $\chi$PT amplitudes can be consistently embedded into a unitary $S$ matrix element.
See for example the discussion in Ref.~\cite{Zhou:2004ms}. }
\begin{itemize}
	\item [1)]  In the region $s\in(c_R,s_R)$, the $S$-matrix $S^{1/2,J}_{\pm}(s)$ must contain a zero\footnote{This argument has been actually used long time ago in Ref.~\cite{PhysRev.123.692}. We thank one referee who brings our attention to this paper.  }, while  $S^{3/2,J}_{\pm}(s)$ does not need to have.
	\item [2)] In the region $s\in(s_L,c_L)$, both  $S^{1/2,J}_{+}(s)$ and $S^{3/2,J}_{-}(s)$   contain a zero for  $J=1/2, 5/2, 9/2,\cdots$, while both $S^{1/2,J}_{-}(s)$ and $S^{3/2,J}_{+}(s)$  contain a zero for  $J=3/2, 7/2, 11/2,\cdots$.
Or in other words, for $I=1/2$ ($I=3/2$), there exists a virtual pole $v_L\in (s_L,c_L)$ in each partial wave amplitude with even (odd) angular momentum.
\end{itemize}
Taking a few examples,  in $S_{31}$ channel there is no virtual pole, while in $p$ waves,
\begin{itemize}
\item $P_{11}$ channel contains a virtual pole  $v_R\in(c_R,s_R)$, as well as the nucleon bound state;
\item $P_{13}$ channel contains a virtual pole in $v_R\in(c_R,s_R)$;
\item $P_{33}$ channel contains a virtual pole in $v_L\in(s_L,c_L)$.
\end{itemize}
A modification of the dispersion representation for function $f$ defined in  Ref.~\cite{Li:2021tnt} in each channel may be needed accordingly, similar to what has been discussed in $S_{11}$ channel~\cite{Li:2021tnt}. Conclusions are similar in all channels: the contribution from the virtual pole to the phase shift is always exactly canceled by  the additional contribution from the dispersion integral in $f$, when $v_L=c_L\, (v_R=c_R)$. Taking for example the case that there exists only one virtual pole $v_L$ on the left. In this situation,
\begin{equation}
S^{cut}=\frac{S^{phy}}{S^{v_L}\prod_iS^i}\ ,
\end{equation}
where $S^{v_L}$ represents the ``$S$ matrix" of the newly found virtual pole $v_L$, while $S_i$ depicts all other known poles.~\footnote{For more details about the production representation, one is refer to Refs.~\cite{Zheng:2003rw}, \cite{Zhou:2006wm},\cite{Yao:2020bxx}.}
Such an $S^{cut}$ is real and negative when $s\in (s_L,c_L)$,~\footnote{$S^{phy}(s_L)=+1$ according to Eq.~(\ref{SM}). This is because $s_L$ is the branch point caused by the $u$ channel continuous absorptive singularities, $T(s_L)$ itself is regular. For  resonances $S^i=+1$, for bound state or virtual state $S^i=-1$, when $s=s_L$. Particularly $S^{v_L}$ is negative when $s\in{(s_L,v_L)}$ and positive when $s\in(v_L,c_L)$.} so the ``spectral function" $f=\ln S^{cut}/2i\rho$ maintains a cut $\in (s_L, c_L)$ and its contribution $\frac{s}{\pi}\int^{c_L}_{s_L}\frac{\mathrm{Im}[\ln S^{cut}(s')/2i\rho(s')]}{s'(s'-s)}ds'$ exactly cancels the contribution from the virtual pole when $v_L=c_L$. In another situation when there exists a virtual pole on the right, i.e., $v_R\in(c_R,s_R)$, the ``spectral function" $f$ is defined the same as in Ref.~\cite{Li:2021tnt}:\footnote{Except $P_{11}$ channel, where there exists a nucleon  bound state,  the contribution of  the virtual state  remains.   }
\begin{equation}\label{overlinef}
	\bar f(s) = \dfrac{\ln-S^{cut}}{2i\rho(s)}- \dfrac{\pi}{2\bar\rho(s)}
\end{equation}
where the   function $\bar\rho(s)$ is the `deformed' $\rho(s)$ with its cut $\in [s_L, s_R]$, while the cut of the latter is defined on $(-\infty, s_L]\cup [s_R,+\infty)$. Further, function $\bar f(s)$ is identical to $f(s)$ when $s$ lies in the physical region. Now $-S^{cut}>0$ when $s\in (s_L,c_L)$ and  $-S^{cut}<0$ when $s\in (c_R,s_R)$. So ${\ln-S^{cut}}/{2i\rho(s)}$ contains a cut $(c_R,s_R)$, on which its imaginary part is canceled by that from the second term on the $r.h.s.$ of Eq.~(\ref{overlinef}). Hence $\bar f$ contains an additional cut $(s_L,c_R)$ coming from
$-{\pi}/{2\bar\rho(s)}$, and its contribution on $(s_L, c_R)$ is exactly canceled by that of the  virtual pole when $v_R=c_R$. As a consequence all calculations made in Ref.~\cite{Wang:2018nwi} remain correct with very high accuracy.

The following discussions dedicate to  determining the location of  those zeros. {Let us focus on $I=1/2$ for the moment.} The  explicit expressions of $S$-matrix elements read
\begin{equation}
		S^{1/2,J}_\pm(s) = \mathcal A^{1/2,J}_{\pm}(s)+\mathcal B^{1/2,J}_{\pm}(s)\ln\dfrac{s-c_L}{s-c_R}\ ,
\end{equation}
with
\begin{align}\label{A1J}
\mathcal A^{1/2,J}_{\pm}(s)=1-\dfrac{im_N^2g^2}{4\pi F^2 s\rho(s)}\left\{(W\pm m_N)(E_N\mp m_N)[\dfrac{P_{J+1/2}(y)}{2}\ln\dfrac{m_N^2}{s}-O_{J-1/2}(y)]+\notag\right.
\\
\phantom{=\;\;}
\left.(W\mp m_N)(E_N\pm m_N)[\dfrac{P_{J-1/2}(y)}{2}\ln\dfrac{m_N^2}{s}-O_{J-3/2}(y)]
\right\}+\cdots\ ,
\end{align}
where $\cdots$ refer to possible but irrelevant regular terms, and
\begin{align}\label{B1J}
\mathcal B^{1/2,J}_{\pm}(s)=-\dfrac{im_N^2g^2}{8\pi F^2 s\rho(s)}[(W\pm m_N)(E_N\mp m_N)P_{J+1/2}(y)+(W\mp m_N)(E_N\pm m_N)P_{J-1/2}(y)]\ ,
\end{align}
where $W\equiv \sqrt{s}$ denotes the  center-of-mass frame energy and $E_N=({s+m_N^2-m_\pi^2})/{2\sqrt{s}}$ is the nucleon energy. It's worth  stressing that the function $\mathcal B^{1/2,J}_{\pm}(s)$  is immune of $\chi$PT corrections.

Let $v^J_{R\pm}\in (c_R,s_R)$ denotes the position of the zero  of $S^{1/2,J}_{\pm}(s)$, which gives
\begin{equation}\label{VR}
S^{1/2,J}_{\pm}(v^J_{R\pm})=\mathcal A^{1/2,J}_{\pm}(v^J_{R\pm})+\mathcal B^{1/2,J}_{\pm}(v^J_{R\pm})\ln\dfrac{v^J_{R\pm}-c_L}{v^J_{R\pm}-c_R}=0\ .
\end{equation}
{Suppose $v^J_{R\pm}$ approaches $c_R$ for  sufficiently large $J$, as will be verified a short while later,} the solution of Eq.~(\ref{VR}) may be  obtained by iteration method,
\begin{equation}\label{VRS}
		v^J_{R\pm}=c_R+(c_R-c_L)e^{\mathcal A^{1/2,J}_{\pm}({c_R})/\mathcal B^{1/2,J}_{\pm}({c_R})}\ .
\end{equation}
Using  Eq.~(\ref{B1J}) and the fact that $P_l[y(c_R)] =P_l(1)=1$, one finds that $\mathcal B^{1/2,J}_{\pm}({c_R})$ is negative and independent of $J$.
Further, when $J$ tends to $\infty$,  $\mathcal A^{1/2,J}_{\pm}({c_R}) $ goes to the infinity. To prove this, notice that the function $O_{l-1}$ defined in Eq.~(\ref{OD}) behaves as (taking $l=2k+1$ for example)
{\begin{equation}
		\begin{split}
&\lim_{k\to\infty}O_{2k}(1)=\lim_{k\to\infty}\sum_{r=0}^{k}\dfrac{(4k-4r+1)}{(2r+1)(2k+1-r)}\\
	&=\lim_{k\to\infty}\sum_{r=0}^{k}\left(\dfrac{2}{2r+1}-\dfrac{1}{2k-r+1}\right)>\lim_{k\to\infty}\sum_{r=0}^{k}\left(\dfrac{2}{2r+1}-\dfrac{1}{k+1}\right)
	\\
	&=\lim_{k\to\infty}\sum_{r=0}^{k}\dfrac{2}{2r+1}-\lim_{k\to\infty}\sum_{r=0}^{k}\dfrac{1}{k+1}\to\infty
	\ .
	\end{split}
\end{equation}}
Hence, according to Eq.~(\ref{A1J}), as $J\to \infty$, the leading contribution of $\mathcal A^{1/2,J}_{\pm}(c_R)$ is\footnote{The contribution from chiral corrections will not diverge, since the contribution itself is regular when $s\in(s_L,s_R)$, and notice that $|P_l(x)|<1$ when $x\in[-1,+1]$ for arbitrary $l$ -- which is proved using the Laplace formula of Legendre function.}
{\be
\mathcal A^{1/2,J}_{\pm}(c_R)\to \dfrac{im_N^2m_\pi^2g^2}{4\pi F^2 c_R\rho(c_R)}{ O_{J-1/2}(1)}\to +\infty\ .
\ee}
From above, one concludes that $v^J_{R\pm}$  gets closer and closer to $c_R$ as $J$ increases, according to Eq.~(\ref{VRS}).
Hence, $c_R$ is the accumulation point of poles on the second sheet. {What we have discussed is actually an example of a general mathematical theorem: \textit{as the sequence is bounded, it must have an accumulation, i.e., it has  one convergent subseuqence.}}
Similarly analyses can be made on the situation in the line segment $(s_L, c_L)$. One finds that both $v_{L+}^{1/2,J}$ and $v_{L-}^{1/2,J}$ approaches $c_L$ when $J\to\infty$.
 Therefore, following the steps of Ref.~\cite{Martin:1970jsp}, we prove that $s=c_L, c_R$ are two accumulation of singularities of $T^{1/2}(s,t)$, on the second sheet of complex $s$-plane. However, the proof in Ref.~\cite{Martin:1970jsp} on the statement that $s=0$ is an accumulation of singularities on the second sheet of $s$ plane for  $\pi\pi$ scattering amplitudes is fully non-perturbative. Our proof given here is valid to all orders of perturbation chiral expansions.

The analyses can also be applied to isospin $3/2$ case, and only final results are presented here:
\begin{itemize}
\item when $J=\frac{4n+1}{2}$ ($n=1,2,3,\cdots$), $S^{3/2,J}_-$ maintains a zero $v^{3/2,J}_{L-}\in(s_L,c_L)$;
\item when $J=\frac{4n-1}{2}$ ($n=1,2,3,\cdots$), $S^{3/2,J}_+$ maintains a zero $v^{3/2,J}_{L+}\in (c_R,s_R)$.
\end{itemize}
One further proves that 
 accumulation of zeros at $c_L$ occur and  that $s=c_L$ is an accumulation of singularities of $T^{3/2}(s,t)$ on the second sheet of $s$ plane.

{\bf Acknowledgement:} The authors would like to thank Wei-Nian Zhang at Sichuan University, and   Wei-Da Zheng for helpful discussions. This work is support in part by National Nature Science Foundations
of China under contract number 11975028 and 10925522.


\begin{thebibliography}{}
\bibitem{Li:2021tnt} Q.~Z.~Li et al., Chin. Phys. {\bf C 46} (2022) 023104.
\bibitem{Kennedy:1962apa} J.~Kennedy and T.~D.~Spearman, Phys. Rev. {\bf 126} (1961) 1596.
\bibitem{Gasser:1988NP}{J.~Gasser, M.~E.~Sainio and A. ~Svarc, Nucl. Phys. {\bf B307}, 779 (1988).}
\bibitem{Zhou:2004ms}Z.~Y.~Zhou et al.,  JHEP {\bf 02} (2005) 043.


\bibitem{PhysRev.123.692}R. Blankenbecler,  M. L. Goldberger,  S. W. MacDowell, S. B. Treiman, Phys. ~Rev.~{\bf 692} (1961) 123.


\bibitem{Zheng:2003rw}H.~Q.~Zheng et al., Nucl.~Phys.~{\bf A 733} (2004) 235.
\bibitem{Zhou:2006wm}Z.~Y. Zhou, H.~Q.~Zheng, Nucl.~Phys.~{\bf A 775} (2006) 212.
\bibitem{Yao:2020bxx}D.~L.~Yao, L.~Y.~Dai, H.~Q.~Zheng, Z.~Y.~Zhou, Rept. Prog. Phys. {\bf 84} (2021) 076201.
\bibitem{Wang:2018nwi}Y.~F.~Wang, D.~L.~Yao, H.~Q.~Zheng, Chin. Phys. {\bf C 43} (2019) 064110 and references therein.
\bibitem{Martin:1970jsp}{A.~Martin, F.~Cheung, \textit{Analyticity Properties and Bounds of the Scattering Amplitudes}. Contribution to: 10th Brandeis University Summer Institute in Theoretical Physics, elementary particle physics and scattering theory (1970).}

\end{thebibliography}
\end{document}